\documentstyle[pre,twocolumn,aps,epsf]{revtex}

\newcommand{\beq}{\begin{equation}}
\newcommand{\eeq}{\end{equation}}
\newcommand{\bea}{\begin{eqnarray}}
\newcommand{\eea}{\end{eqnarray}}
\newcommand{\req}[1]{Eq.~(\ref{#1})}
\newcommand{\gcc}{{\rm~g\,cm}^{-3}}
\newcommand{\ApJ}[1]{Astrophys.~J. {\bf #1}}

\newcommand{\PRL}[1]{Phys.\ Rev.\ Lett. {\bf #1}}
\newcommand{\PR}[2]{Phys.\ Rev.\ #1 {\bf #2}}

\begin{document}
\draft
\title{Equation of state of fully ionized electron-ion plasmas.\\
II. Extension to relativistic densities and to the solid phase}

\author{
Alexander Y.~Potekhin$^{1,}$\thanks{Electronic address:
palex@astro.ioffe.rssi.ru}
 and 
Gilles Chabrier$^{2}$ 
}
\address{$^1$Ioffe Physical-Technical Institute,
     194021 St.\ Petersburg, Russia}
\address{$^2$%
     Ecole Normale Sup\'erieure de Lyon,
       CRAL (UMR CNRS No.\ 5574),
     69364 Lyon Cedex 07, France}

\date{Phys. Rev. E {\bf 62} (2000) 8554--8563}
\maketitle

\begin{abstract}
The analytic equation of state of nonideal Coulomb plasmas
consisting of pointlike ions immersed in a polarizable
electron background
[G.~Chabrier and A.~Y. Potekhin, \PR{E}{58},  4941 (1998)]
is improved, and its applicability range is considerably extended.
First, the fit of the electron screening contribution
in the free energy of the Coulomb liquid
is refined at high densities 
where the electrons are relativistic.
Second, we calculate the screening contribution for
the Coulomb solid (bcc and fcc)
and derive an analytic fitting expression.
Third, we propose a simple
approximation to the internal and free energy
of the liquid one-component plasma of ions,
accurate within the numerical errors of the most recent
Monte Carlo simulations.
We obtain an updated value of the coupling parameter at
the solid-liquid phase transition for the one-component plasma:
$\Gamma_m=175.0\pm0.4\,(1\sigma)$.
\end{abstract}


\section{Introduction}
Fully ionized electron-ion plasmas (EIP)
are encountered in laboratory experiments, in stellar and planetary
interiors, in supernova explosions, etc.
From the theoretical point of view, 
the free energy of fully ionized EIP provides 
the reference system for 
models aimed at describing the thermodynamic properties 
of partially ionized plasmas. 
Thus the studies of EIP are of both theoretical and practical interest.

In a previous paper \cite{CP98} we have 
calculated thermodynamic quantities of Coulomb plasmas
consisting of pointlike ions immersed in a compressible,
polarizable electron background
and devised analytic fitting formulae for these quantities.
The calculations were based on a linear-response theory 
for the ion-electron ($ie$) interaction, which is valid as long as 
the typical $ie$ interaction energy $(Ze)^2/2a_0$
(where $a_0$ is the Bohr radius, and $Ze$ is the ion charge)
is smaller than the kinetic energy of the electrons. 
This condition is fulfilled either at temperatures $T\gtrsim10^5Z^2$~K
or at densities $\rho\gtrsim AZ^2\gcc$,
where $A$ is the ion mass number.
For the nonrelativistic regime, i.e., at densities $\rho\ll10^6\gcc$, finite-temperature effects were included 
in the electronic dielectric function,
as well as the local-field correction
arising from electron correlation effects,
following the model developed in
Ref.\cite{C90}. In the relativistic regime,
similar calculations were done using the Jancovici \cite{Janco}
dielectric function.

Since the electron screening is weak at high densities,
and since the bulk of calculations have been performed
using the nonrelativistic model,
our fit for the 
$ie$ contribution was not very accurate at $\rho\gtrsim10^6\gcc$,
where the electrons are relativistic.
Because of the same weakness of the screening,
this inaccuracy in the $ie$ contribution at high $\rho$
did not deteriorate the overall
accuracy for the {\em excess\/} part of the 
free energy, which sums up the ion-ion ($ii$),
electron-electron ($ee$), and $ie$ contributions.
There are, however, physical problems which require 
an accurate evaluation of
the $ie$ part even at high densities
(an example is mentioned below).
In this paper we present a modification
of the analytic formula \cite{CP98} for the $ie$ free energy
which improves significantly the accuracy in the domain of 
relativistic electrons,
keeping unchanged the previous nonrelativistic results.

Second,
we calculate the $ie$ part of the free energy for a Coulomb
solid, where the ions form either a body-centered-cubic (bcc)
or face-centered-cubic (fcc) lattice.
The calculation is performed in a perturbation 
approximation, which is accurate because 
the screening is weak. We employ an analytic 
expression for the ion structure factor $S(k)$
of a Coulomb crystal, obtained in Ref.\ \cite{BKPY} 
in the harmonic approximation 
for large wave numbers $k$ outside the first Brillouin zone.
For small $k$, we supplement it by an exact limiting
form of $S(k)$.
We evaluate the screening contribution for both the classical
and quantum harmonic crystals
and construct a fitting formula which accurately reproduces
our numerical results.

The above mentioned improvements of the equation of state
are significant at densities $\rho\gtrsim10^6\gcc$. 
Such densities cannot be reached in the laboratory,
but they are commonly encountered in the interiors of white dwarfs
and envelopes of neutron stars (e.g., Ref.~\cite{ShapiroTeukolsky}).

In addition, we present simple formulae
for the excess internal and free energies
of a classical one-component plasma (OCP) liquid,
which take into account the most recent Monte Carlo (MC)
results \cite{DWS,Caillol}, and which are accurate for any values
of the Coulomb coupling parameter from the gaseous phase
to the dense liquid regime.
Analyzing various results for the free energy 
of the OCP liquid and solid,
we revise the value of the coupling parameter at
the solid-liquid phase transition.

In the next section, we describe the basic parameters of the EIP.
In Sec.~\ref{sect-ii}, we consider 
the OCP liquid and determine its freezing point.
In Sec.~\ref{sect-ie}, we present an improved fit
to the free-energy contribution 
due to the electron screening in a Coulomb liquid.
In Sec.~\ref{sect-ie-solid}, we evaluate an analogous contribution
for a Coulomb solid and fit it by an analytic expression.

\section{Thermodynamic parameters}
\label{sect-param}
We consider EIP consisting of pointlike 
ions and electrons.
The basic dimensionless parameters are 
the electron density parameter $r_s$
and the ion coupling parameter $\Gamma$:
\beq 
   r_s={a_e\over a_0},
\quad
    \Gamma = {(Ze)^2\over k_B T a} = \Gamma_e Z^{5/3}, 
\quad  
    \Gamma_e = {e^2\over k_B T a_e}.
\eeq
Here, $k_B$ is the Boltzmann constant,
$a_e=(\frac43\pi n_e)^{-1/3}$
is the mean inter-electron distance, 
$a=(\frac43\pi n_i)^{-1/3}=a_e\,Z^{1/3}$
is the mean inter-ion distance,
and $n_e$ ($n_i$) denotes the electron (ion) number density.
$\Gamma_e$ has a meaning of the
coupling parameter for nondegenerate electrons.

Quantization of the ionic motion
is important if 
$
   T\ll T_p = \hbar\omega_p/k_B,
$
where $\omega_p = (4\pi Z^2e^2 n_i/m_i)^{1/2}$
is the ion plasma frequency, $m_i$ being the ion mass.
A corresponding dimensionless parameter is
\beq
   \eta = T_p/T=\Gamma\,\sqrt{3/R_S},
\eeq
where 
\beq
   R_S={a m_i\over\hbar^2}\, (Ze)^2 = {m_i\over m_e}\,r_s\,Z^{7/3}
\eeq 
is the ion density parameter.
We neglect ion quantum-exchange
effects, which is justified if $R_S\gg\Gamma$
(see, e.g., Ref.\ \cite{JonesCeperley}).

The electrons are characterized by the degeneracy parameter $\theta$ 
and the relativity parameter $x_r$,
\beq
    \theta = T/ T_F,
\qquad 
    x_r = p_F/(m_e c),
\eeq
where $T_F$ 
is the Fermi temperature,
$c$ is the speed of light, and $p_F=\hbar\,(3\pi^2 n_e)^{1/3}$ 
is the Fermi momentum. 
The electron screening properties are determined by the Thomas-Fermi
wave number
\beq
   k_{\rm TF} = \left( 4\pi e^2 {\partial n_e / \partial \mu} \right)^{1/2},
\label{k_TF}
\eeq
where $\mu$ is the electron chemical potential.

For these parameters,
the following estimates are accurate within 0.005\%:
\bea
&&   x_r \approx 0.014005\, r_s^{-1}\approx
       1.0088 \left(\rho_6\,Z/A\right)^{1/3},
\\&&
   \Gamma_e \approx {22.547\, x_r \over T_6},
\quad 
   \theta^{-1}\approx {5930\over T_6}\left(\sqrt{1+x_r^2}-1\right),
\eea
where $\rho_6=\rho/(10^6\gcc)$ and $T_6=T/(10^6$~K).
In the nonrelativistic plasma ($x_r\ll1$), 
$\theta\approx 0.543 \,r_s/\Gamma_e$.
In the ultrarelativistic case ($x_r\gg1$),
$\theta\approx (263\,\Gamma_e)^{-1}$.
If the electrons are nondegenerate ($\theta\gg1$),
$k_{\rm TF}a_e\approx \sqrt{3\Gamma_e}$.
For strongly degenerate electrons ($\theta\ll1$),
\beq
    k_{\rm TF}a_e\approx0.185\,(1+x_r^{-2})^{1/4}.
\eeq
The ion quantum parameter $\eta$ is expressed through $x_r$ and $\Gamma$
as
\beq
   \eta\approx 0.3428\,\Gamma\,\sqrt{x_r}\,Z^{-7/6}A^{-1/2}.
\label{eta-est}
\eeq

Within the aforementioned approximation of weak electron-ion coupling,
the total Helmholtz free energy $F_{\rm tot}$
can be written as
\beq
   F_{\rm tot} = 
   F_{\rm id}^{(i)} + F_{\rm id}^{(e)} 
   + F_{ee} + F_{ii} + F_{ie}, 
\label{f-tot}
\eeq
where $F_{\rm id}^{(i,e)}$ 
denote the ideal free energy
of ions and electrons, respectively, 
and the last three terms
represent an excess free energy arising from 
interactions.
$F_{\rm id}^{(i)}$ is the free energy of an ideal Boltzmann gas.
For the electrons at arbitrary degeneracy and relativism,
$F_{\rm id}^{(e)}$ can be expressed through
Fermi-Dirac integrals and approximated by 
analytic formulae \cite{CP98}.
An analytic parameterization for
the nonideal (exchange and correlation)
part of the free energy
of the nonrelativistic electrons, $F_{ee}^{\rm nr}$, has been given
in Ref.~\cite{IIT}.
For the relativistic electrons, the exchange free energy $F_x^{\rm rel}$
has been given, e.g., in Ref.\ \cite{SB96}, while
the correlation corrections are negligible
because they contain an additional small factor
$\sim\alpha_f\ln|\alpha_f|$ \cite{YS},
where $\alpha_f\approx1/137$ is the fine-structure constant.
In practice, we use the following interpolation
between the nonrelativistic and relativistic regimes:
if $\Gamma_e\geq0.07$ and $r_s\leq0.13$, we set
\bea
   F_{ee}&=&(1-\xi)\,F_{ee}^{\rm nr}+\xi\,F_x^{\rm rel},
\\
   \xi&=&\exp[-(\Gamma_e/0.07-0.9)^{-2}-(0.13/r_s-0.9)^{-2}];
\nonumber
\eea
otherwise we set $F_{ee}=F_{ee}^{\rm nr}$.
The interpolation is sufficiently smooth, because
$F_{ee}^{\rm nr}$ and $F_x^{\rm rel}$
closely match each other
at the chosen boundary between the two regimes. 

In the following sections we consider
the last two terms in \req{f-tot},
which represent the excess free energy
of an OCP of ions and the contribution due to the
ion-electron interactions, respectively.

\section{OCP and melting transition}
\label{sect-ii}
Liquid and solid phases of the OCP have been 
studied extensively 
by various analytic and numerical methods.
All the thermodynamic functions of the classical OCP
can be expressed as functions of the only parameter
$\Gamma$. 
At $\Gamma\ll1$, a diagrammatic cluster expansion yields
\bea
   u_{ii} &\equiv& {U_{ii}\over N_ik_B T} =
            -{\sqrt{3}\over2}\,\Gamma^{3/2}
            -3\Gamma^3\,\left[\frac38\ln(3\Gamma)
            +{C_E\over2}-\frac13\right]
 \nonumber\\&&
            -\Gamma^{9/2}(1.6875\,\sqrt{3}\,\ln\Gamma - 0.23511)
            +\cdots,
\label{abe}
\eea
where $C_E=0.57721\ldots$ is the Euler constant.
Here, the first term is the Debye-H\"uckel energy,
the second one is due to Abe \cite{Abe}, and the $\,\sim\Gamma^{9/2}$ one
is due to Cohen and Murphy \cite{CohenMurphy}.
Since $F_{ii}$ vanishes at high $T$, 
it can be obtained from $U_{ii}$ by integration:
\bea
   f_{ii} &\equiv& {F_{ii}\over N_ik_B T} =
     \int_0^\Gamma {u_{ii}(\Gamma')\over\Gamma'} {\rm\,d}\Gamma'
\label{int}
\\
      &=& -{ \Gamma^{3/2}\over\sqrt3}
            -\Gamma^3 \left(\frac38\,\ln\Gamma+0.24225\right)
 \nonumber\\&&
            -\Gamma^{9/2} (0.64952\,\ln\Gamma - 0.19658 )
            +\cdots.
\label{fren_abe}
\eea

The above analytic expansion is not applicable
for $\Gamma\gtrsim1$. 
The most accurate up to date numerical results 
for the internal energy of the liquid OCP at $1\leq\Gamma\leq200$
have been obtained by MC
simulations by DeWitt and Slattery \cite{DWS} and by Caillol \cite{Caillol}.
These authors have also constructed analytic fits to their data
with the standard deviations
comparable to the numerical MC noise.
Unfortunately, these fits cannot be extended to small $\Gamma$, 
which hampers obtaining the free energy by \req{int}.
On the other hand, the hypernetted-chain (HNC)
result for $F_{ii}$ at $\Gamma=1$
is slightly inaccurate because the HNC approximation
neglects the so called bridge functions in the
diagrammatic representation of the interactions.
To circumvent the difficulty, 
DeWitt and Slattery \cite{DW} used small differences between HNC and MC 
at $\Gamma = 0.8$ and 0.6 to get the
corrected value of $f_{ii}(\Gamma = 1) = -0.4368$.

We propose a different approach. 
We consider the parameterization
\beq
 u_{ii} = \Gamma^{3/2}
   \left[{A_1\over\sqrt{\Gamma+A_2}} 
   + {A_3\over\Gamma+1}\right]
      + {B_1\,\Gamma^2 \over\Gamma+B_2}
      + {B_3\,\Gamma^2 \over\Gamma^2+B_4},
\label{fitionu}
\eeq
where $A_3=-\sqrt{3}/2-A_1/\sqrt{A_2}$.
The terms in square brackets have been used in Ref.~\cite{CP98},
the term with $B_1$ provides an adjustment of the fit
to the MC data at large $\Gamma$,
and the last term adjusts to \req{abe} at small $\Gamma$.
The best-fit parameters with respect to the data \cite{DWS,Caillol}
are given in Table~\ref{tab-ii}.
Then the free energy can be obtained from \req{int}:
\bea
&&
  f_{ii} = A_1\left[\sqrt{\Gamma(A_2+\Gamma)}
           -A_2\,\ln\!\left(\sqrt{\Gamma/ A_2}
           +\sqrt{1+\Gamma/ A_2}\right)\right]
 \nonumber\\&& \quad
          + 2A_3\left[\sqrt{\Gamma}
           -\arctan\sqrt{\Gamma} \right]
\nonumber\\&& \quad
         +B_1\left[\Gamma 
         - B_2\ln\left(1+{\Gamma\over B_2}\right)\right]
           +{B_3\over2}\,\ln\left(1+{\Gamma^2\over B_4}\right).
\label{fition}
\eea
The corresponding expression for heat capacity is
\bea
   {C_{V,ii}\over N_i k_B} &=& {\Gamma^{3/2}\over2}\left[
      A_3\,{\Gamma-1\over(\Gamma+1)^2}-{A_1\,A_2\over(\Gamma+A_2)^{3/2}}
       \right]
\nonumber\\&&
    + \Gamma^2\left[B_3\,{\Gamma^2-B_4\over(\Gamma^2+B_4)^2}
        - {B_1\,B_2\over(\Gamma+B_2)^2}\right].
\label{cvliq}
\eea 

\begin{figure}
    \begin{center}
    \leavevmode
    \epsfxsize=86mm
    \epsfbox[110 170 470 680]{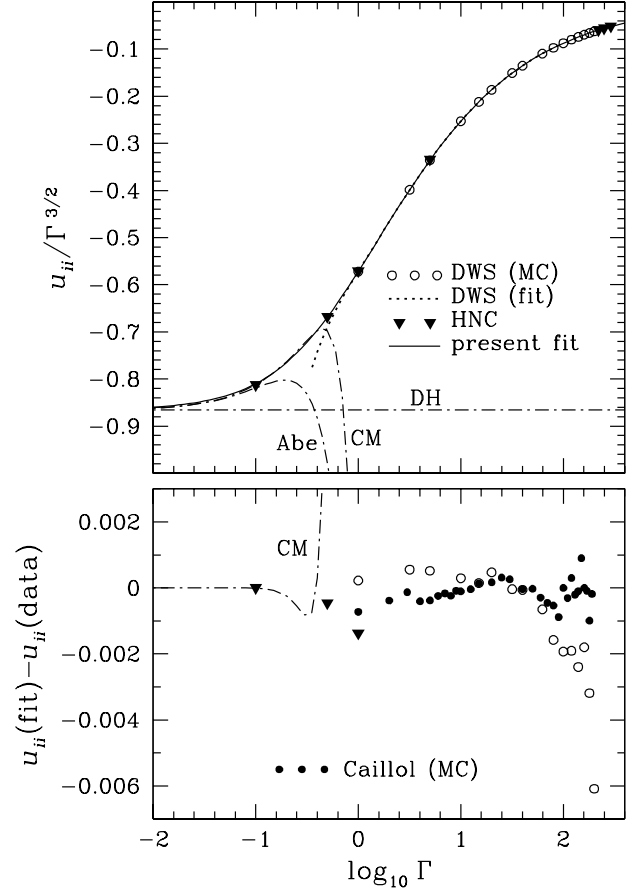} 
    \end{center}
\caption{
Upper panel: comparison of the fit (\protect\ref{fitionu}) (solid line) 
with the Debye-H\"uckel (DH), Abe~\protect\cite{Abe}, 
and Cohen-Murphy~\protect\cite{CohenMurphy} (CM)
approximations (dot-dashed lines),
with the MC results (circles) and the fit (dotted line) of Ref.~\protect\cite{DWS} (DWS),
and with some of our HNC results (triangles).
Lower panel: residual differences between the fit (\protect\ref{fitionu})
and (i) the analytic expansion (\protect\ref{abe}) (dot-dashed line),
(ii) results of HNC calculations (triangles),
(iii) MC results of Ref.~\protect\cite{DWS} (open circles),
and (iv) numerical results of Ref.~\protect\cite{Caillol} (MC+extrapolation).
}
\label{fig-ii}
\end{figure}

Comparison of \req{fitionu} with \req{abe}
at $\Gamma<1$ and with the MC data from Refs.~\cite{DWS,Caillol}
at $\Gamma\geq1$, supplemented by some of our HNC calculations, 
is given in Fig.~\ref{fig-ii}.
The upper panel displays the ratio $u_{ii}/\Gamma^{3/2}$
(which is constant in the Debye-H\"uckel approximation).
The magnitude of the possible error is demonstrated by
the lower panel. Here, 
the dot-dashed line shows the difference between 
the approximation (\ref{fitionu}) 
with the second set of parameters
and expansion (\ref{abe}), while
various symbols show residual differences between 
the same approximation and numerical 
(HNC and MC) results.
The distribution of the residuals around zero
looks irregular, which indicates that they represent
 a numerical noise of the MC calculations rather than
 an error of the fit (\ref{fitionu}).
In addition, we have checked that the difference between our fit to 
the free energy, \req{fition},
and the one in Ref.\ \cite{DWS} (at $\Gamma\geq1$) is of the order of
the aforementioned small uncertainty in
$f_{ii}(\Gamma=1)$.

More complicated interpolations between the low- and high-$\Gamma$ 
limits were proposed previously \cite{Kahlbaum,SB99}. 
By construction,
they reproduce exactly \req{abe} at $\Gamma\to0$
and the fits to MC results at $\Gamma\gg1$. 
Compared with the present fit, however, those interpolations
have somewhat larger differences 
from the HNC results at $0.1<\Gamma<1$.


\begin{figure}
    \begin{center}
    \leavevmode
    \epsfxsize=86mm
    \epsfbox[120 155 480 330]{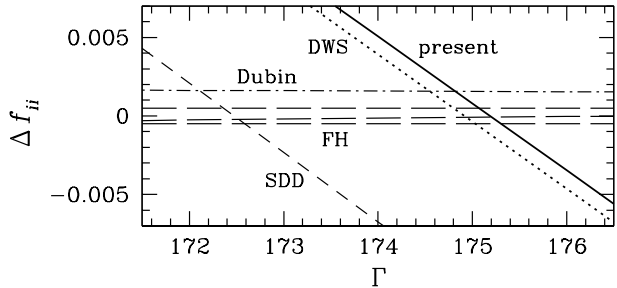} 
    \end{center}
\caption{ 
Difference between the free energy of the solid OCP given
by a 3-parameter fit of Ref.\ \protect\cite{FH} and parameterizations
for the liquid OCP according to Refs.\ \protect\cite{SDD}
(SDD; short-dashed line), 
\protect\cite{DWS} (dotted line), \protect\req{fition} (solid line),
and for the solid according to Dubin
\protect\cite{Dubin} (dot-dashed line).
The long-dashed
lines marked ``FH'' correspond to the 
4-parameter fit and to the $\pm1\sigma$-uncertainty
of the 3-parameter fit in Ref.~\protect\cite{FH}.
}
\label{fig-fliqsol}
\end{figure}

The freezing of Coulomb OCP liquid 
into a bcc crystal occurs 
when the free energy of the solid becomes 
lower than that of the liquid at $\Gamma=\Gamma_m$.
Nagara et al.\ \cite{NNN} and Dubin \cite{Dubin},
having improved a previous treatment of 
anharmonic corrections to the free energy of the Coulomb crystal,
obtained $\Gamma_m=172\pm1$.
However, these authors employed
an older fit \cite{SDD} (SDD) for the liquid. 
Figure~\ref{fig-fliqsol} shows the {\em differences}
between $f_{ii}$ for the solid and liquid OCP
given by various parameterizations.
For the solid, we have adopted 
the three-parameter fit by Farouki and Hamaguchi \cite{FH}
to their molecular dynamics simulations in the range
$170\leq\Gamma\leq400$.
The horizontal long-dashed lines correspond to
the standard deviation of that fit. 
The line between them represents a four-parameter fit \cite{FH}
in the same $\Gamma$ interval.
The dot-dashed line shows the difference between the fit 
of Ref.\cite{FH} and that by Dubin \cite{Dubin}.
The value of $\Gamma_m$ indicated above
is given by the intersection of the latter line with
the short-dashed one (SDD).
Using updated results for the OCP liquid
(either Ref.~\protect\cite{DWS} or our \req{fition}, 
represented by the dotted and solid line, respectively)
and the OCP solid \cite{FH}, we obtain $\Gamma_m=175.0\pm0.4$.

\section{Electron screening in a Coulomb liquid}
\label{sect-ie}
We now consider electron polarization effects in the EIP.
In the previous paper \cite{CP98}, we have calculated
$F_{ie}$ using the model developed in Ref.\cite{C90}
for nonrelativistic EIP.
The HNC equations have been
solved numerically for an
effective screened inter-ion potential $V_{\rm eff}$,
which is the sum of the bare ionic potential 
and the induced polarization
potential, to obtain $F_{ii} + F_{ie}$ and corresponding contributions
to the internal energy ($U_{ii} + U_{ie}$) 
and pressure ($P_{ii} + P_{ie}$).
The same equations solved for the bare Coulomb potential
give $F_{ii}$, $U_{ii}$, and $P_{ii}$.
The difference represents the screening ($ie$) part.
Inclusion of the finite-temperature effects in $V_{\rm eff}$
provides a correct treatment
of the thermodynamic quantities over a wide range of values of $\Gamma$
from the Debye-H\"uckel limit $\Gamma\ll1$
to the strong-coupling limit $\Gamma\gg1$ for various $r_s$ and $Z$.

Relativistic calculations have been performed
employing the same HNC technique but with the Jancovici \cite{Janco}
dielectric function $\varepsilon(k,x_r)$,
which is appropriate at strong degeneracy ($\theta\ll1$) 
and arbitrary $x_r$. 
The results are in good agreement
with those obtained by Yakovlev and Shalybkov \cite{YS},
who have used an equation
\beq
   f_{ie}\equiv {F_{ie}\over N_i k_B T}
       = -{\Gamma a\over\pi} \int_0^\infty S(k,\Gamma) 
     \, [\varepsilon(k,x_r)-1]\, dk,
\label{Galam}
\eeq
where $S(k,\Gamma)$ is the static structure factor of ions
(i.e., the Fourier transform
of the ion radial distribution function).
Equation (\ref{Galam})
has been derived by Galam and Hansen \cite{GalamHansen}
using a thermodynamic perturbation
scheme, which can be represented as an expansion
in powers of $k_{\rm TF}$.
We have repeated the calculations \cite{YS}
using a more recent and accurate $S(k,\Gamma)$ \cite{YCD} than in the
original work; the change in $f_{ie}$
due to this update does not exceed 4\%.

Note that \req{Galam} differs from the standard
first-order perturbation approximation by a replacement
of $[1-1/\varepsilon]$ by $[\varepsilon-1]$. 
The resulting difference $\approx (k_{\rm TF}a)^3/6$
has the same order of magnitude as the second-order 
perturbation correction \cite{GalamHansen}.
Our HNC calculations with the Jancovici dielectric function
at $x_r\lesssim1$ and $\Gamma\geq1$ 
coincide within 2\%
with the results of Ref.\ \cite{YS}, 
whereas the substitution of $[1-1/\varepsilon]$ in \req{Galam}
yields a considerable difference:
for example, for $\Gamma=1$ and $Z=26$ this difference
approaches 40\% even at large $x_r$.
We conclude that the approximation (\ref{Galam}) 
is very accurate at high densities.

The screening contribution to the free energy of the Coulomb liquid
at $0 < r_s\lesssim1$, $0<\Gamma\lesssim300$, and $1\leq Z \leq 26$
has been fitted by the expression \cite{CP98}
\beq
  f_{ie} \equiv {F_{ie}\over N_i k_B T} = -\Gamma_e \,
   { c_{\rm DH} \sqrt{\Gamma_e}+
    c_{\rm TF} a \Gamma_e^\nu g_1 h_1
    \over
     1+\left[ b\,\sqrt{\Gamma_e}+ a g_2 \Gamma_e^\nu/r_s \right]
    h_2 },
\label{fitscr}
\eeq
where 
$
   c_{\rm DH} = (Z/\sqrt{3})
    \left[(1+Z)^{3/2}-1-Z^{3/2}\right]
$
ensures exact transition to the Debye-H\"uckel limit at $\Gamma\to0$,
$c_{\rm TF} = 
   (18/175)\,(12/\pi)^{2/3} 
   Z^{7/3}\left(1-Z^{-1/3}+0.2\,Z^{-1/2}\right)
$
reproduces the Thomas-Fermi limit \cite{Salpeter} 
at $Z\to\infty$,
the parameters
$   a = 1.11 \, Z^{0.475}$,
$
   b = 0.2+0.078 \,(\ln Z)^2,
$ and
$
   \nu = 1.16 + 0.08 \ln Z
$
provide a low-order approximation to $F_{ie}$ for intermediate
$r_s$ and $\Gamma$, and the functions
\bea
   g_1 &=& 1 + 0.78 \left[
           21+ \Gamma_e(Z/r_s)^3 \right]^{-1} (\Gamma_e/Z)^{1/2},
\nonumber\\
   g_2 &=& 1+{ Z-1 \over 9}
         \left(1+\frac{1}{0.001\, Z^2+2\Gamma_e}\right)
         {r_s^3 \over 1+ 6 \, r_s^{2} }
\nonumber
\eea
improve the fit at relatively large $r_s$.
The results of our nonrelativistic finite-temperature HNC calculations
are reproduced by setting $h_1$ and $h_2$ equal to unity;
these factors come into play in the relativistic case.


\begin{figure}
    \begin{center}
    \leavevmode
    \epsfxsize=86mm
    \epsfbox[86 170 512 650]{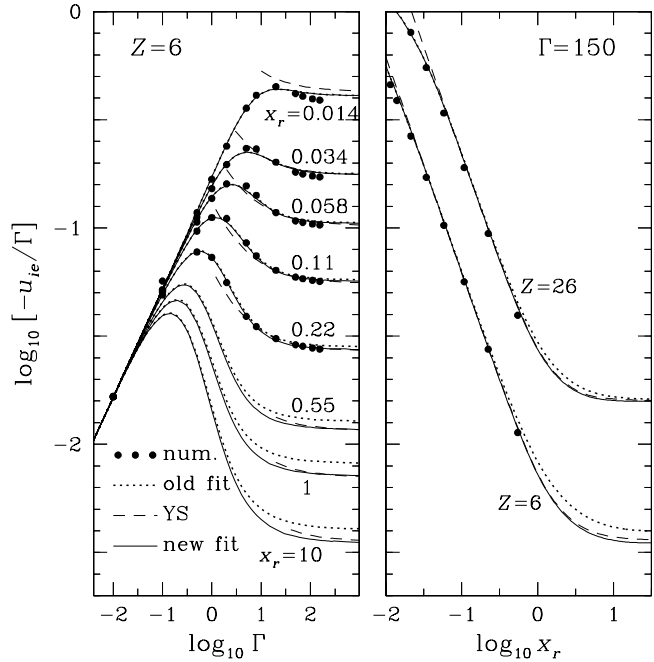} 
    \end{center}
\caption{
Calculated (filled circles) and fitted (solid lines)
normalized contribution to the internal energy due to polarization,
$u_{ie}$, as function of $\Gamma$ at different values of 
$x_r$, $Z=6$ (left panel) and as function of $x_r$ 
at two values of $Z$, $\Gamma=150$ (right panel). 
For comparison, approximations \protect\cite{CP98} (dotted lines)
and \protect\cite{YS} (dashed lines) are also shown.
}
\label{fig-ie}
\end{figure}

In the latter case, the asymptotic behavior of \req{fitscr}
at $\Gamma\to\infty$
should change from $f_{ie}\propto \Gamma r_s$ 
to $f_{ie}\propto \Gamma r_s \sqrt{1+x_r^2}$.
This is achieved simply by setting $h_2=(1+x_r^2)^{-1/2}$.
Then the zero-temperature
Thomas-Fermi limit \cite{Salpeter}
($r_s\ll1$, $\Gamma\to\infty$, $Z\to\infty$)
is reproduced exactly.

The factor $h_1$ is devised to correct
the fit at finite $Z$ in the relativistic domain.
A form chosen previously \cite{CP98}
was not very accurate,
as illustrated by the dotted lines
in Fig.~\ref{fig-ie} for the internal energy
\beq
   u_{ie} \equiv {U_{ie}/( N_i k_B T)} = 
         \partial f_{ie}(r_s,\Gamma)/\partial\ln\Gamma.
\label{u_ie}
\eeq
A more accurate relativistic correction reads
\beq
  h_1(x_r) = {1+ x_r^2 / 5 \over
      1+0.18\,Z^{-1/4} x_r + 0.37\, Z^{-1/2} x_r^2 +x_r^2 / 5}.
\eeq
The resulting $u_{ie}$ [\req{u_ie}]
is plotted by the solid lines in Fig.~\ref{fig-ie}.
There is now a good agreement with the thermodynamic perturbation
expansion \cite{YS} at large $\Gamma$ for any $x_r$, without
deteriorating the accuracy of 
the old fit in the nonrelativistic domain.
Quantitatively, for $1 < \Gamma < 100$ and $x_r < 0.25$,
the difference between the fit and the HNC results is
typically 2--3\%,
with maximum 8\% for $Z=1$, $\Gamma=100$ and $r_s=2.074$
(the maximum $r_s$ value used in the calculations).
Note that the model of EIP has only marginal physical relevance
at such large values of $r_s$ and $\Gamma$
because of the incipient bound-state formation.
On the other hand, at very strong coupling
($\Gamma\geq100$) and relativistic densities ($x_r>0.1$),
the results of Ref.\ \cite{YS} and of our relativistic HNC
calculations are reproduced
by our fit with typical deviation of 1--3\%
(maximum 4.3\% at $Z=6$, $\Gamma=100$, and $x_r=10$).

\begin{figure}
    \begin{center}
    \leavevmode
    \epsfxsize=86mm
    \epsfbox[70 270 515 570]{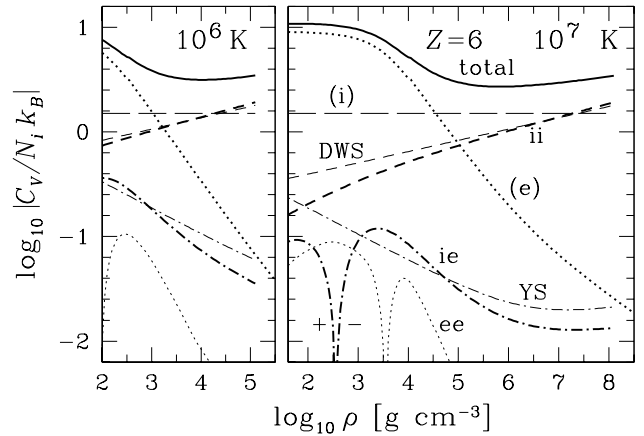} 
    \end{center}
\caption{
Absolute values of the
heat capacity of fully ionized liquid carbon at $T=10^6$~K
and $10^7$~K. Dotted curves show the contributions of
the electrons (heavy line -- ideal Fermi-gas contribution,
light line -- exchange and correlation correction),
dashed lines -- contributions of the ions 
(long dashes -- ideal-gas part, short dashes -- correlation part),
and dot-dashed curves -- ion-electron (polarization) correction.
The latter curves end at $\Gamma=175$.
The dips on the $ee$ and $ie$ curves signify a change
of sign. For the $ii$ and $ie$ contributions,
present approximations (heavy lines) are compared with those
in Refs.\protect\cite{DWS} (DWS) and \protect\cite{YS} (YS)
(light lines).The heavy solid line shows the sum of all terms.
}
\label{fig-cvliq}
\end{figure}

The heat capacity per ion in units of $k_B$, $C_V/N_i k_B$, 
of the classical EIP liquid is shown in Fig.~\ref{fig-cvliq}
for $Z=6$, $T=10^6$~K and $10^7$~K. These plasma conditions can occur,
for example, in interiors of some giant stars
or in accreted envelopes of neutron stars \cite{PCY}.
Various contributions, shown in the figure,
correspond to separate terms in \req{f-tot}.
At relatively low densities, the main contribution is that of the electrons,
with the limiting value $\frac32Z=9$.
With increasing $\rho$, the electron gas becomes degenerate,
and its heat capacity decreases.
Then $C_V$ is determined by the ion liquid.
The Coulomb ($ii$) contribution slightly
exceeds the kinetic one ($\frac32$) near freezing.
According to the equipartition theorem,
in a classical ionic crystal the potential 
and kinetic contributions are each equal to $\frac32$ 
(apart from small anharmonic corrections).
This means that freezing is accompanied by a drop
of $C_V$, equal to the excess of the potential
contribution over the kinetic one in the ionic 
liquid just before freezing.
We see however that this excess (and hence the drop)
is not large.

The values of $C_{V,ii}$ determined by \req{cvliq} (thick dashes)
and derived from the fit in Ref.\cite{DWS} (thin dashes)
are close to each other near the freezing. With decreasing density, however,
a large difference develops, which is natural because
the formula in Ref.\cite{DWS} is not applicable at small $\Gamma$.
Of the same origin is the striking discrepancy
between the approximations
for $C_{V,ie}$ derived from \req{fitscr} (thick dot-dashed curve)
and from the fit in Ref.~\cite{YS} (YS), seen at low $\rho$.
In this domain, our fit describes the change
of sign of  $C_{V,ie}$ from negative in the strong-coupling
regime to positive in the Debye-H\"uckel domain.
However, an appreciable difference with Ref.~\cite{YS}
persists even at large $\rho$, where both fits
describe $f_{ie}$ equally 
well (within uncertainties in the structure factor).
This reflects insufficient accuracy
of the present-day determination of the functional form
of $S(k,\Gamma)$ for the strongly coupled Coulomb liquid.

\section{Electron screening in a Coulomb solid}
\label{sect-ie-solid}
\subsection{Perturbation approximation}
\label{sec-pert}
At high densities and below a certain temperature,
the ionic Coulomb plasma forms a Wigner crystal.
For example, interiors of cool white dwarfs \cite{WD}
are expected to be in the solid state.
The cooling 
is governed essentially by the compressibility
and heat capacity of their interiors,
whose central regions are compressed to 
relativistic densities.
In that case, the main contributions to the internal energy
(the zero-temperature electron-gas kinetic energy
and the ion electrostatic part) do 
not depend on temperature, so that
the heat capacity is entirely determined by small
temperature-dependent corrections.
Therefore, evaluation of the polarization corrections
for the Coulomb solid is important for astrophysical applications.

Since the maximum ion frequency in the solid is quite small compared to
the electron plasma frequency, one can use the adiabatic
(i.e., Born-Oppenheimer) approximation,
which allows to decouple the electron and ion dynamics.
Even so, a calculation of the  thermodynamic functions of a
Coulomb solid with allowance for the $ie$ interactions
is a complex problem. A rigorous treatment would consist in 
calculating the dynamical matrix
and solving a corresponding dispersion 
relation for the phonon spectrum.
The first-order
perturbation approximation for the dynamical matrix 
of a classical Coulomb solid with the polarization corrections,
based on an {\em effective\/} inter-ion potential,
was derived by Pollock and Hansen 
\cite{PollockHansen}.
In a quantum crystal, strictly speaking, one would have to
consider the electron-phonon interactions, in order to calculate
the perturbed spectrum.

As mentioned above, the polarization of the electron gas is weak
at the high densities we are interested in.
This suggests a simpler, semiclassical perturbation
approach to evaluate the polarization corrections.
The ionic crystal without $ie$ interactions
is a natural reference model.
Note that the effective inter-ion potential 
in the adiabatic perturbation approximation \cite{PollockHansen}
is just the electrostatic potential, common to the liquid and solid
phases. The difference of this potential from
the bare Coulomb potential can be considered as perturbation.
Then we can apply the Galam-Hansen \cite{GalamHansen}
perturbation theory, which is based on the {\em exact\/}
expression for the free energy involving an integration
over a coupling parameter related to the ``strength'' 
of the perturbation. 
Thus we recover \req{Galam} in the case of solid,
with $S(k)$ replaced by 
$(4\pi)^{-1} \int S({\bf k}) d\Omega$,
where $d\Omega$ is a solid angle element in the direction of 
${\bf k}$.

The resulting polarization correction (\ref{Galam})
does not take into account quantum aspects of the
$ie$ (electron-phonon) interactions, but
it allows us to study effects
arising from quantum modifications 
of the ion-ion correlations. These correlations are
described by
the structure factor $S$, which depends in this case 
on ${\bf k}$, $\Gamma$, and $\eta$.

\subsection{Structure factor}
In a crystal, 
the static structure factor is given by
\beq
     S({\bf k},\Gamma,\eta) = {1\over N_i} \sum_{i,j}
     e^{i {\bf k}\cdot ({\bf R}_i - {\bf R}_j)}
     \left\langle e^{i {\bf k}\cdot \hat{\bf u}_i}
     e^{-i {\bf k}\cdot \hat{\bf u}_j} \right\rangle_T,
\label{Sqt-det}
\eeq
where $\hat{\bf u}_i$
is an operator of ion displacement from an equilibrium
lattice position ${\bf R}_i$, and $\langle\ldots\rangle_T$
denotes the canonical average.
The structure factor (\ref{Sqt-det}) can be decomposed
into elastic (or static-lattice) and inelastic parts,
\beq
    S({\bf k},\Gamma,\eta)=S'({\bf k},\Gamma,\eta)+S''({\bf k},\Gamma,\eta).
\label{separation}
\eeq
The elastic part
is (e.g., Ref.~\cite{Kittel})
\beq
    S_{\rm sol}'({\bf k},\Gamma,\eta) =
(2\pi)^3 n_i\,
 e^{-2W({\bf k},\Gamma,\eta)} {\sum_{\bf G}}' \delta({\bf k}-{\bf G}),
\label{S'}
\eeq
where 
$\sum_{\bf G}'$ denotes a summation over all
reciprocal lattice vectors
${\bf G}$ but ${\bf G}=0$, and
$e^{-2W} \equiv 
   \left\langle \exp(i{\bf k}\cdot\hat{\bf u}) \right\rangle_T^2$
is the Debye-Waller factor.
In isotropic (e.g., cubic) crystals,
one has 
\beq
   2\,W({\bf k},\Gamma,\eta)=r_T^2(\Gamma,\eta)\, k^2 / 3,
\eeq 
where $r_T^2 = \langle \hat{\bf u}^2\rangle_T$ 
is the mean-squared ion displacement (cf.\cite{Kittel}).
In a harmonic crystal, 
\beq
r_T^2 = {a^2\eta\over\Gamma}
        \left[ {\mu_{-1}\over2} + \left\langle 
{\omega_p\over\omega_\nu}\,{1\over \exp(\hbar\omega_\nu/k_BT)-1} 
      \right\rangle_{\rm ph}
          \right],
\label{rT2a}
\eeq
where $\nu \equiv ({\bf q},s)$,
$s=1,2,3$ enumerates phonon modes,
${\bf q}$ is a phonon wave vector,
$\omega_\nu$ is the frequency, 
$\langle\ldots\rangle_{\rm ph}$
denotes averaging over phonon wave vectors and polarizations,
and $\mu_n \equiv \langle (\omega_\nu/\omega_p)^n \rangle_{\rm ph}$.
In the classical limit ($\eta\to0$), $r_T^2=\mu_{-2} a^2/\Gamma$;
and in the quantum limit ($\eta\to\infty$), 
$r_T^2\sim\mu_{-1}a^2\eta/(2\Gamma)$.
Numerical values of $\mu_{-1}$ and $\mu_{-2}$ 
are given in Table~\ref{tab-lattice}.
At arbitrary $\eta$,
a convenient analytic approximation to $r_T^2$ 
is provided by a model of the harmonic Coulomb crystal \cite{CAD} 
which treats two 
acoustic modes as degenerate Debye modes with 
$\omega_\nu=\alpha\,\omega_p\, q/q_{\rm BZ}$,
where $q_{\rm BZ}=(6\pi^2 n_i)^{1/3}$
is the equivalent radius of the Brillouin zone,
and the longitudinal mode as an Einstein mode
with $\omega_\nu=\gamma\omega_p$.
Accuracy of this model for 
the thermodynamics of the bcc Coulomb crystal
has been demonstrated in Ref.\cite{C93},
where the values $\alpha=0.399$ and $\gamma=0.899$
have been derived from the requirement that the model should 
reproduce the exact values of $\mu_{-2}$ and $\mu_2=\frac13$.
For the fcc lattice 
we obtain $\alpha=0.413$ and $\gamma=0.892$.
Using this model, we can calculate the second term
in \req{rT2a}, which yields
\beq
   r_T^2={a^2\over\Gamma} \left[ {\mu_{-1}\eta\over2}
       + {\eta\over3\gamma\,(e^{\gamma\eta}-1)} + {2\over\alpha^3\eta}\,
      \int_0^{\alpha\eta} {t\,dt\over e^t-1} \right].
\label{rT2}
\eeq
This approximation ensures the correct classical and quantum limits.
Between these limits,
the maximum deviation from accurate numerical results \cite{Baiko}
reaches 1.6\% at $\eta\approx9$ for both bcc and fcc lattices.

According to \req{separation},
\req{Galam} can be rewritten as
\bea
     f_{ie} &=& f_{ie}'+f_{ie}'',
\label{f_ie_sum}
\\
     f_{ie}' &=& - {3 \Gamma \over 2}
             {\sum_{\bf G}}' {\varepsilon(G,x_r)-1 \over (G a)^2}
         \exp[-2W(G,\Gamma,\eta)],
\\
    f_{ie}'' &=& -{\Gamma a\over\pi} \int_0^\infty S''(k,\Gamma,\eta) 
      \, [\varepsilon(k,x_r)-1]\, dk.
\label{Galam''}
\eea

The inelastic part of the structure factor of a harmonic crystal
reads \cite{BKPY}
\bea
   S'' &=& e^{-2W} \sum_{\bf R} e^{i{\bf k}\cdot{\bf R}}
         \left[ e^{v({\bf R})k^2}-1\right],
\label{S''exact}
\\
    v(R) &=& {3\hbar\over 2m_i}\left\langle 
        {({\bf k}\cdot{\bf e}_\nu)^2 \over k^2}\,
        {\cos({\bf q}\cdot{\bf R}) \over 
          \omega_\nu \tanh(\hbar\omega_\nu/2k_B T)} 
           \right\rangle_{\rm ph}
\label{v(R)}
\eea
(where ${\bf e}_\nu$ is a phonon polarization vector).
A straightforward use of this expression
is impractical because of a slow convergence of the sum.
For this reason, we employ the approximation \cite{BKPY}:
\beq
   S''(k,\Gamma,\eta) \approx S_0''(k,\Gamma,\eta)  
      \equiv 1-\exp[-2W(k,\Gamma,\eta)].
\label{S0''}
\eeq
As argued in Ref.\ \cite{BKPY}, this approximation is good
for use in integrals over ${\bf k}$ at $k>q_{\rm BZ}$.
In papers addressed to
the transport properties of Coulomb plasmas \cite{BKPY,PBHY}, 
integrals over $k$
were truncated from below at $k=q_{\rm BZ}$.
In \req{Galam''}, however, it is essential to recover
the correct limiting behavior of $S''(k)$ at $k\to0$,
since $[\varepsilon-1]\propto k^{-2}$ becomes large in this limit.
Therefore we use a piecewise approximation:
\beq
    S''=\left\{\begin{array}{ll}
       S_0'' & (k\geq k_1),
\\
    S_1''
  \equiv \frac12 k^2\,[\partial^2 S''/\partial k^2]_{k\to0}
                & (k<k_1),   \end{array}  \right.
\label{piecewise}
\eeq
where the parameter $k_1$ will be determined below.
The exact result for classical Coulomb plasmas
\cite{VH75} reads $S_1''(k) = (ka)^2/(3\Gamma)$.
In general case, $S_1''(k)$ can be found from \req{S''exact}.
At small $k$, the expression in the square brackets in 
\req{S''exact} can be replaced by $v({\bf R})k^2$,
which corresponds to the one-phonon approximation.
Changing the order of averaging and summation,
we see that the summation yields delta function
$\delta({\bf k}\pm{\bf q}-{\bf G})$;
therefore ${\bf q}=\pm{\bf k}$ 
as long as $k < \min G \approx 2q_{\rm BZ}$.
Hence, 
only the longitudinal phonon mode contributes in this limit.
The frequency of this mode in a Coulomb crystal 
tends to $\omega_p$ at small $q$ (e.g., \cite{PollockHansen}),
which enables us to perform averaging in \req{v(R)}.
Finally we obtain
\beq
   S_1''(k,\Gamma,\eta) = {(ka)^2\over 6\Gamma}\,{\eta\over\tanh(\eta/2)}.
\label{S1''}
\eeq

In order to test our approximation
(\ref{piecewise}) and to find the optimum value of $k_1$,
let us consider the electrostatic energy $U_{\text{el-st}}$
of a Coulomb crystal,
\beq
   u_{\text{el-st}} \equiv {U_{\text{el-st}}\over N_i k_B T}
     = {\Gamma a\over\pi} \int_0^\infty [ S(k)-1 ]\,dk
      = u'+u'',
\eeq
where, according to Eqs.~(\ref{separation}) and (\ref{S'}),
\beq
   u' = {3\Gamma\over2}\, {\sum_{\bf G}}'\, 
     {e^{-2W(G,\Gamma,\eta)}\over(Ga)^2}
\label{u'}
\eeq
is the static-lattice part.
Baiko et al.\cite{Baiko99} have shown that
\[
{u'\over\Gamma } = -C_M + {r_T^2\over 2a^2} 
      + \sqrt{3\over\pi}\,{a\over 2r_T} + \cdots,
\]
where the terms not explicitly written are exponentially small
at large $\Gamma$.
For the inelastic contribution, our model yields
$u'' = u_0'' + u_1''$, where
\bea
   u_0'' &=& {\Gamma a\over\pi} \int_0^\infty [ S_0''(k)-1 ]\,dk
     = - \sqrt{3\over\pi}\,{\Gamma a\over 2r_T},
\\
  u_1'' &=& {\Gamma a\over\pi} \int_0^{k_1} [ S_1''(k)-S_0''(k) ]\,dk
\nonumber\\
     &=& {(k_1 a)^3\over18\pi} {\eta\over\tanh(\eta/2)}
     - {\Gamma a\over\pi} \!\left[k_1 
  - {\sqrt{3\pi} \over 2 \, r_T}\,{\rm erf}{r_T k_1\over\sqrt{3}}
       \right].
\label{u1''}
\eea

\begin{figure}
    \begin{center}
    \leavevmode
    \epsfysize=70mm
    \epsfbox[120 270 450 570]{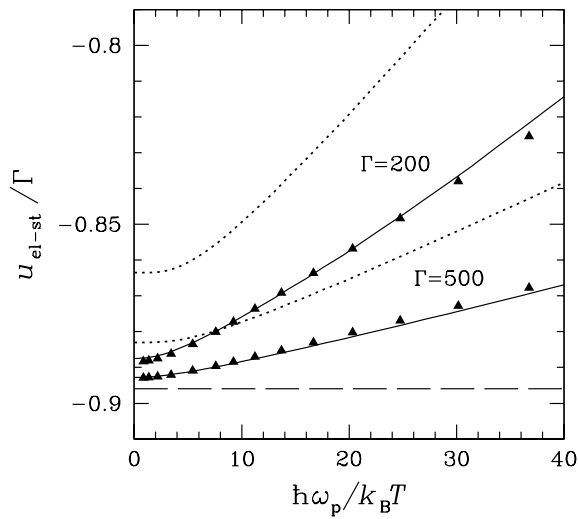} 
    \end{center}
\caption{
Normalized electrostatic energy of a bcc Coulomb crystal
calculated using the approximate structure factor 
given by \protect\req{piecewise} (solid lines) 
compared with analogous calculations
with a model structure factor, \protect\req{S0''} (dotted lines),
and with accurate numerical calculations \protect\cite{Baiko}
(triangles). Upper curve of each type or symbol corresponds to $\Gamma=200$
and lower one to $\Gamma=500$.
Long-dashed line displays the Madelung limit.
}
\label{fig-uelst}
\end{figure}

On the other hand, in the harmonic lattice approximation,
$
  U_{\text{el-st}} = - N_i C_M (Ze)^2/a + U_v/2,
$
where the first term represents the energy of 
a perfect ionic lattice in uniform electron background,
$C_M$ being the Madelung constant (Table~\ref{tab-lattice}),
and, from the virial theorem,
the second term is one half of the vibrational energy
of a harmonic crystal, 
\beq
   U_v = 3\,N_i k_B T \left[ \left\langle{\omega_\nu\over\omega_p}\,
   {\eta\over e^{\eta\omega_\nu/\omega_p}-1}\right\rangle_{\rm ph}
   + {\mu_1\,\eta\over2} \right].
\eeq
We determine $k_1$ so as to recover 
the classical limit 
$u_{\text{el-st}}=-C_M\Gamma + 3/2$
at $\eta=0$ and $\Gamma\to\infty$.
This yields
\beq
   {k_1\over q_{\rm BZ}} = 
   \left({\mu_{-2}-3 \over \mu_{-2}-1}\right)^{1/3} \approx 0.94.
\eeq
Figure~\ref{fig-uelst} shows $u_{\text{el-st}}$ 
calculated from Eqs.\ (\ref{u'})--(\ref{u1''})
for the bcc crystal at {\em finite\/} $\eta$ and $\Gamma$
(dot-dashed lines),
compared with a calculation in which $S''$ is set equal to $S_0''$ 
at any $k$ (dotted lines) 
and with results of numerical calculations \cite{Baiko}. 
We see that our modification of
the structure factor at $k<0.94\, q_{\rm BZ}$ provides a 
significant improvement 
over the model without such modification
(denoted as HL1 in Ref.\cite{Baiko99}).

\subsection{Results}
Using Eqs.\ (\ref{f_ie_sum})--(\ref{Galam''}) 
and (\ref{S0''})--(\ref{S1''}), 
we have calculated the polarization correction
$f_{ie}$ over a wide range of parameters:
$80\leq\Gamma\leq3\times10^4$,
$10^{-2}\leq x_r\leq10^2$,
and $1\leq Z \leq 92$.
Not all combinations of the considered parameters
are physically relevant; for instance,
at $Z=1$ and large $x_r$ the ion-exchange effects
neglected in our study become important.
The use of such extended set of parameters, however,
delivers robustness to a fitting formula presented
below.

\begin{figure}
    \begin{center}
    \leavevmode
    \epsfxsize=86mm
    \epsfbox[60 180 550 640]{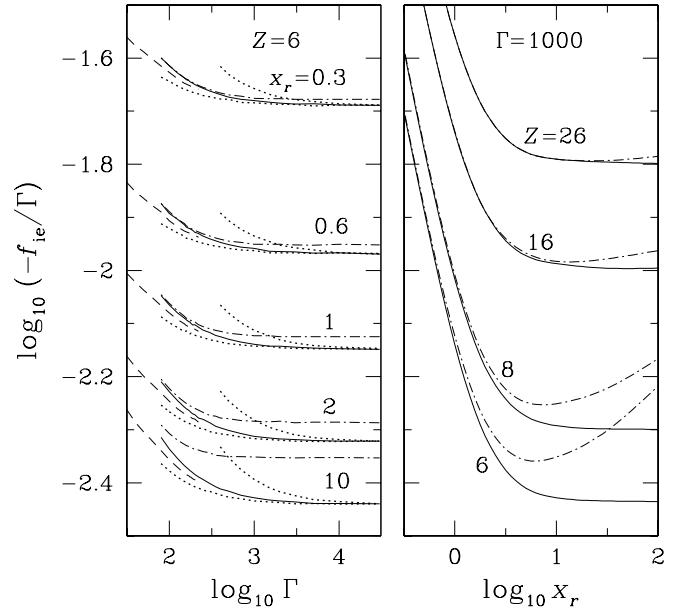} 
    \end{center}
\caption{
Normalized polarization correction $f_{ie}$ 
to the free energy of a Coulomb crystal.
Left panel: $\log_{10}(-f_{ie}/\Gamma)$ against $\log_{10}\Gamma$
at $Z=6$ for several values of $x_r$.
Right panel: $\log_{10}(-f_{ie}/\Gamma)$ against $\log_{10}x_r$
at $\Gamma=10^3$ for several values of $Z$.
Solid lines: a classical solid; dot-dashed lines: 
quantum effects included.
On the left panel, dotted lines show the results
for the classical solid with simplified structure factors,
and dashed lines show the results in a liquid.
}
\label{fig-f-ie-sol}
\end{figure}

Some results of our calculations are shown in Fig.~\ref{fig-f-ie-sol}.
Solid lines correspond to the piecewise approximation (\ref{piecewise})
of the structure factor
in the classical case ($\eta\to0$).
Dashed lines on the left panel
reproduce calculations in the liquid with 
$S(k,\Gamma)$ from Ref.\cite{YCD}. 
The upper and lower dotted lines at every value of $x_r$
show, respectively, the results of calculations 
with the inelastic part of the structure factor replaced by $S_0''$ 
(as in the HL1 model of Ref.\cite{Baiko99})
and by 0 (as in Refs.\cite{BKPY,PBHY}).
Compared to these simplified approximations, 
the present model provides
a smaller discontinuity of $f_{ie}$ at the freezing point
(near the ends of the dashed lines).
On the other hand, the divergence of the dotted curves
towards smaller $\Gamma$ shows 
that the result is still model-dependent.
This model dependence disappears
at $\Gamma\gtrsim3000$, since the
static-lattice contribution becomes relatively large.

In reality, at large values of $\Gamma$ and small values of $x_r$
shown in Fig.~\ref{fig-f-ie-sol}, the quantization of ionic vibrations
becomes important. This quantization considerably modifies
the structure factor.
This effect is taken into account by letting $\eta$ to be finite
in Eqs.~(\ref{rT2}) and (\ref{S1''}).
Results of the calculations, where $\eta$ was determined
from \req{eta-est} assuming $A=2Z$, 
are plotted in Fig.~\ref{fig-f-ie-sol} by dot-dashed lines.
The curves on the left panel become flat as $\eta$ becomes large,
which corresponds to 
an approximate proportionality $f_{ie}\propto\Gamma$.
As a consequence, the polarization contribution
to the specific heat, $C_{V,ie}$, goes to zero at large $\eta$
(but remains one of the leading contributions, as shown below).

The numerical results can be fitted by the expression
\beq
   f_{ie} = -f_\infty(x_r)\,\Gamma
         \left[ 1 + A(x_r)\,(Q(\eta)/\Gamma)^s \right],
\label{fitscr-sol}
\eeq
where
\bea
   f_\infty(x) &=& a_{\rm TF} Z^{2/3} b_1\,\sqrt{1+b_2/x^2},
\nonumber\\
   A(x) &=& { b_3+a_3 x^2 \over 1+b_4 x^2 },
\nonumber\\
   Q(\eta) &=& \sqrt{1+(q\eta)^2},
\nonumber
\eea
and parameters $s$ and $b_1$--$b_4$ depend on $Z$:
\bea
   s &=& \left[ 1+0.01\,(\ln Z)^{3/2} + 0.097\,Z^{-2} \right]^{-1},
\nonumber\\
  b_1 &=& 1 - a_1 \,Z^{-0.267} + 0.27\,Z^{-1},
\nonumber\\
  b_2 &=& 1 + {2.25\over Z^{1/3}}\,
      {1+a_2\,Z^5+0.222\,Z^6 \over 1+0.222\,Z^6},
\nonumber\\
  b_3 &=& a_4/(1+\ln Z),
\nonumber\\
  b_4 &=& 0.395 \ln Z + 0.347\, Z^{-3/2}.
\nonumber
\eea
The parameter $a_{\rm TF}$, related to $c_{\rm TF}$ in \req{fitscr},
is chosen so as to reproduce the Thomas-Fermi limit \cite{Salpeter} 
at $Z\to\infty$: 
$a_{\rm TF}=(54/175)(12/\pi)^{1/3}\,\alpha_f = 0.00352$.
The numerical parameters $a_1$--$a_4$ and $q$ are slightly
different for bcc and fcc crystals;
they are given in Table~\ref{tab2}.

For a classical crystal, an average error of the fit is 1\%
for all $Z$, $x_r$, and $\Gamma$,
and the maximum error is 3.1\% at $Z=92$, $\Gamma\gtrsim10^4$,
and $x_r\approx2$.
In the quantum case ($\eta\neq0$), the fit is accurate
for $Z\geq3$ only. In the range $3\leq Z\leq 30$,
an average error is 1\%, and a maximum 3\% 
occurs at $Z=3$,
$\Gamma\approx100$, and $x_r\approx2$.

\subsection{Discussion}
The results presented in Fig.~\ref{fig-f-ie-sol} indicate that,
although the polarization 
in a Coulomb crystal is very weak, it does not vanish
even at arbitrarily large $\Gamma$ and $x_r$.
As in the case of strongly coupled liquid, 
$f_{ie}$ is roughly proportional to $(k_{\rm TF}a)^2$, which tends
to a finite limit at relativistic densities.
The order of magnitude of 
the screening correction $\delta_{ie}=F_{ie}/F_{\rm tot}$
for a classical Coulomb plasma
at arbitrarily high densities
is given by the Thomas-Fermi result \cite{Salpeter},
which is reproduced by \req{fitscr-sol} at 
$\Gamma\to\infty$ and $Z\to\infty$:
$\delta_{ie}\approx0.004\,Z^{2/3}$.
Quantitative difference of the perturbation result
at finite $Z$
from the Thomas-Fermi limit is quite noticeable, $\sim Z^{-0.3}$.

\begin{figure}
    \begin{center}
    \leavevmode
    \epsfxsize=86mm
    \epsfbox[25 170 550 650]{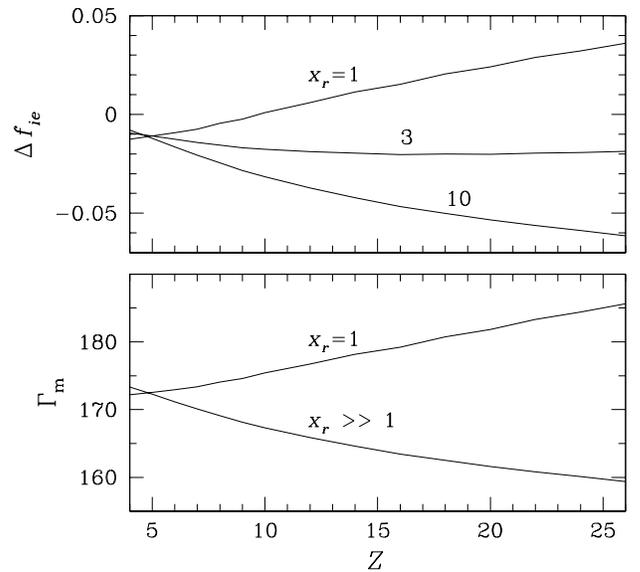} 
    \end{center}
\caption{
Upper panel: difference between polarization corrections
to the free energy in the solid and liquid phases 
at $\Gamma=175$, $x_r=1$, 3, 10.
Lower panel: Coulomb coupling parameter at the melting point
when polarization corrections are taken into account.
}
\label{fig-fliqsolz}
\end{figure}

As mentioned in Sec.~\ref{sec-pert}, 
our treatment of the screening
contribution is approximate.
Nevertheless, we can use these results in order to demonstrate
the importance of the polarization corrections.
On the upper panel of Fig.~\ref{fig-fliqsolz},
the difference $\Delta f_{ie}$ between $f_{ie}$ values in the 
solid and liquid Coulomb plasmas
at the OCP melting point $\Gamma=175$
is plotted against $Z$ for three values of $x_r$.
The largest $x_r=10$ represents virtually
the ultrarelativistic limit.
When compared to Fig.~\ref{fig-fliqsol},
this plot shows that $\Delta f_{ie}$
is sufficiently large to affect $\Gamma_m$.
This effect is shown on the lower panel of Fig.~\ref{fig-fliqsolz},
where we have plotted our estimate of $\Gamma_m$ at $x_r=1$ and $x_r\gg1$.
Since $\Delta f_{ie}$ remains finite at any $x_r$,
the classical OCP value $\Gamma_m=175$ is never
exactly recovered even at arbitrarily large $\rho$.

Another important effect of the polarization corrections
in the solid phase is that on the specific heat $C_V$.
By differentiation of \req{fitscr-sol}, we obtain
\bea
 u_{ie} &=& - f_\infty \Gamma \left[ 1+A\,(1-s/Q^2)\,(Q/\Gamma)^s\right],
\\
 {C_{V,ie}\over N_i k_B} & = &
      f_\infty\, s A \left({\Gamma\over Q} \right)^{1-s}
        { (q\eta)^2-1+s \over Q^3 }.
\eea
In a classical crystal, $C_{V,ie}$ is only a small negative
correction to the total $C_V\approx3N_ik_B$.
When $T$ decreases much below $T_p$, the heat capacity of 
an ionic crystal \cite{C93} goes to zero as
$C_{V,i}\approx 1.6N_ik_B\pi^4/(\alpha\eta)^3\propto T^3$,
whereas the $ie$ contribution becomes positive and decreases
as 
\beq
   C_{V,ie}\sim
    N_i k_B f_\infty\, s A\, (R_S/3q^2)^{(1-s)/2} (q\eta)^{-1}
       \propto T,
\label{cv-ie-sol}
\eeq
at the same rate as the heat capacity
of a strongly degenerate electron gas \cite{YS},
\beq
C_{V,e}\sim 
Z N_i \,(k_B T /m_e c^2)\,\pi^2\,\sqrt{1+x_r^2}/x_r^2.
\eeq
Equation (\ref{cv-ie-sol}), derived from the fit (\ref{fitscr-sol}),
agrees with the limiting expression at $\eta\to\infty$
which follows from Eqs.~(\ref{Galam}),
(\ref{S'}), (\ref{S0''}), and (\ref{rT2}):
\bea
  {C_{V,ie}\over N_ik_B} &\sim& {\pi^2\over \alpha^3\eta}\,
 \left[{\sum_{\bf G}}' \,{\varepsilon(G,x_r)-1\over3}
\right.\nonumber\\&&\left.
      -{2a^3\over9\pi}\int_{k_1}^\infty[\varepsilon(k,x_r)-1]\,k^2\,dk\right].
\label{cv-lim}
\eea
Thus 
$C_{V,ie}$ becomes larger than $C_{V,i}$ at sufficiently low $T$,
which probably signifies that the thermodynamic perturbation
theory is violated at this $T$.

The discussed effect is of anharmonic nature.
Indeed, the harmonic approximation for the Hamiltonian
leads to the Debye law $C_V\propto T^3$, regardless of inclusion
of the polarization correction in the force matrix.
Therefore the dependence $C_{V,ie}\propto T$ in 
Eqs.~(\ref{cv-ie-sol}) and (\ref{cv-lim}) is due to the use of the 
full Coulomb potential (not only its harmonic part) in the 
$ie$ interaction energy, which has led to \req{Galam}.

It is also noteworthy that the modification of the OCP
structure factor by the quantum effects 
renders $C_{V,ie}$ positive. 
A plain extrapolation of the $ie$ contribution from the 
liquid regime into the solid would be completely inappropriate,
as it would result in a negative total heat capacity.

\begin{figure}
    \begin{center}
    \leavevmode
    \epsfxsize=86mm
    \epsfbox[90 230 490 600]{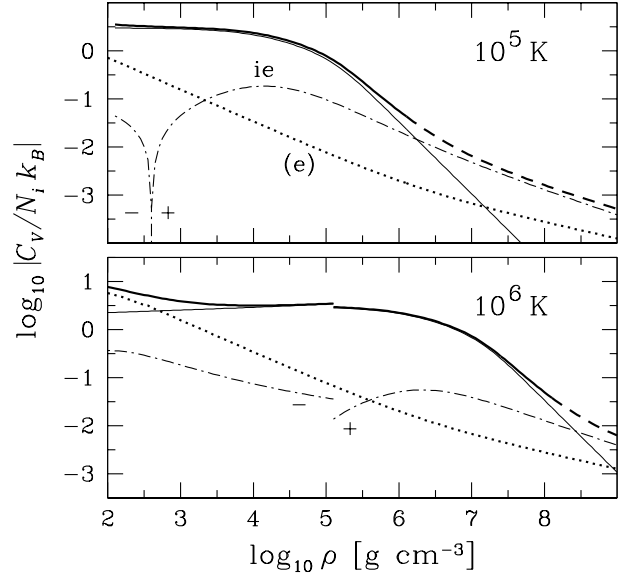} 
    \end{center}
\caption{
Absolute values of heat capacity of carbon at high densities
for two values of $T$. The contributions
of free electrons, ionic OCP, and electron-ion
interaction are shown by dotted, thin solid,
and dot-dashed lines, respectively.
The thick curve shows the total value.
The dashed part of the latter curve corresponds to the region
where the thermodynamic perturbation theory used in the calculation
of $C_{V,ie}$ is not reliable.
}
\label{fig-cv06sol}
\end{figure}

The behavior of different contributions to the heat capacity
in the solid phase as function of $\rho$ and $T$
is illustrated in Fig.~\ref{fig-cv06sol}.
Here we consider $^{12}$C at $10^5$~K
and $10^6$~K. In the latter case (the bottom panel)
one can see also the discontinuities
at the liquid-solid phase transition at $\rho\approx10^5\gcc$,
discussed above. 
As in a liquid, we can safely neglect 
the exchange correction, which at $x_r\gg1$ is as small as
$-(\alpha_f/2\pi)\,C_{V,e}\sim -10^{-3}C_{V,e}$.
At relatively low densities, $C_V$ is determined mainly by the ionic
contribution.
As $T_p$ becomes greater than $T$ with increasing density,
the phonon contribution to $C_V$ freezes out rapidly,
and $C_V$ becomes determined by the degenerate electron gas,
polarized by the electric field of ions.

This may have important consequences for astrophysical applications.
In particular, the heat capacity of old white dwarfs,
whose temperature is so low that their interiors
are formed of quantum Coulomb crystals \cite{WD}, may be 
substantially influenced by the polarization effects \cite{Chabrier00}.

\section{Summary}
We have improved analytic approximations \cite{CP98} 
for the contributions to the free energy of a Coulomb 
liquid due to the $ii$ and $ie$ correlations.
In addition, we have 
suggested an approximation for the $ie$ contribution
to the free energy of a Coulomb crystal.

An improvement of the $ii$ part enables us to determine accurately
the classical OCP melting point.
The Coulomb coupling parameter at the phase transition
is found to be $\Gamma_m\approx175$,
slightly larger than a previously determined value.
An improvement of the $ie$ part in the liquid phase
yields a better precision at densities
$\rho\gtrsim10^6\gcc$, where the electrons are relativistic.
Finally, our estimates of the $ie$ part
of the free energy of a Coulomb crystal
show that it is important for 
applications. For example,
our results demonstrate that it affects the melting
of a classical Coulomb crystal 
and may contribute appreciably 
to the heat capacity of a quantum crystal.
Since our calculations for the Coulomb solid
are based on an approximate method
and performed using an approximate structure factor,
the latter results can be considered as estimates only.
These estimates show, however,
that the polarization corrections in Coulomb crystals
are not as unimportant as it was often believed;
they deserve to be studied further using more elaborate methods.

\begin{acknowledgements}
We thank F.~Douchin and L.~Segretain for useful discussions,
H.\,E. DeWitt and T.~Kahlbaum for helpful communications,
and D.\,A. Baiko and D.\,G. Yakovlev for critical remarks.
A.Y.P.\ is grateful to the theoretical astrophysics group
at the Ecole Normale Sup\'erieure de Lyon for hospitality 
and financial support.
The work of A.Y.P.\ has been partially
supported by INTAS Grant No.\ 96-542
and RFBR Grant No.\ 99-02-18099.
\end{acknowledgements}



\begin{table}
\caption{Parameters of \protect\req{fitionu}.\protect\tablenote{Powers of 10 are given in square brackets.} 
}
\label{tab-ii}
\begin{tabular}{ccccccc}
data & $-A_1$ & $A_2$ & $B_1$ & $B_2$ & $-B_3$ & $B_4$ \\
\noalign{\smallskip}\hline\noalign{\smallskip}
 Ref.\cite{DWS} & $0.9070\phantom{00}$ & $0.62954$ 
       & $4.56[-3]$ & $211.6$ & $1.0[-4]$ & $4.62[-3]$ \\
 Ref.\cite{Caillol} & $0.907347$ & $0.62849$ 
       & $4.50[-3]$ & $170.0$ & $8.4[-5]$ & $3.70[-3]$ \\
\end{tabular}
\end{table}

\begin{table}
\caption{Parameters of Coulomb crystals \protect\cite{Baiko}.}
\label{tab-lattice}
\begin{tabular}{ccccc}
  lattice type & $\mu_{-2}$ & $\mu_{-1}$ & $\mu_1$ & $C_M$ \\
\noalign{\smallskip}
\hline
\noalign{\smallskip}
 bcc & 12.973 & 2.798\,55 & 0.511\,3875 & 0.895\,929\,255\,682 \\
 fcc & 12.143 & 2.719\,82 & 0.513\,1940 & 0.895\,873\,615\,195 \\
\end{tabular}
\end{table}

\begin{table}
\caption{Parameters of \protect\req{fitscr-sol}.}
\label{tab2}
\begin{tabular}{cccccc}
  lattice type & $a_1$ & $a_2$ & $a_3$ & $a_4$ & $q$ \\
\noalign{\smallskip}
\hline
\noalign{\smallskip}
 bcc & 1.1866 & 0.684 & 17.9 & 41.5 & 0.205 \\
 fcc & 1.1857 & 0.663 & 17.1 & 40.0 & 0.212 \\
\end{tabular}
\end{table}

\end{document}